\documentclass[aps,prd,reprint]{revtex4-1}
\usepackage{amsmath,amssymb,amsthm,graphicx}
\usepackage[mathscr]{eucal}
\usepackage[colorlinks,linkcolor=blue,
anchorcolor=blue,citecolor=green]{hyperref}
\usepackage[dvipsnames,usenames]{color}

\usepackage{physics}
\usepackage{lipsum}
\usepackage{epsf,epsfig,graphics,soul}
\usepackage{verbatim,color,ulem,booktabs}

\begin{document}
\title{Acoustic quasibound states and tachyonic instabilities from binary
Bose-Einstein condensates}
\date{\today}
\author{Wei-Can Syu}
\email{syuweican@gmail.com}
\affiliation{Department of
	Physics, National Dong-Hwa University, Hualien 974301, Taiwan, R.O.C.}
\author{Da-Shin Lee}
\email{dslee@gms.ndhu.edu.tw}
\affiliation{Department of
	Physics, National Dong-Hwa University, Hualien 974301, Taiwan, R.O.C.}

\begin{abstract}
We consider two-component Bose-Einstein condensates (BECs) and introduce the BEC vortex in $1+2$ dimensions.
 We focus on two types of gapped excitations induced by  the modes of two-component BECs with relative phases of $0$ and $\pi$, analogous to the massive scalar field with positive and negative mass squared, respectively. 
 The inclusion of space-dependent Rabi coupling can induce an effective space-dependent mass term. We study superradiant instabilities resulting from the quasibound states corresponding to positive mass squared and the tachyonic instabilities arising from negative mass squared in both the frequency and time domains. 
 These instabilities resemble two possible mechanisms that make Kerr black holes unstable in scalar-tensor gravity with the presence of matter around the black hole. Our proposed  phenomena could potentially be implemented in future experiments, drawing from the success of recent analog rotating black hole implementations.
\end{abstract}

\keywords{Bose-Einstein condensate, analogue Gravity, localization}
\pacs{04.70.Dy, %Quantum aspects of black holes, evaporation, thermodynamics
04.62.+v, %Quantum fields in curved spacetime
03.75.Kk. %Dynamic properties of condensates; collective and hydrodynamic excitations, superfluid flow
}
\maketitle
\newpage
\section{INTRODUCTION}\label{sec_introduction}
The linear perturbations around a perturbed black hole can be probed through its damped oscillatory behavior in the form of so-called quasinormal modes (QNMs) \cite{Berti2009,Sam2012,Konoplya2011}. The QNMs have a complex frequency, where the real part gives the frequency of oscillations and the imaginary part determines the lifetime. During the ringdown phase of a perturbed black hole, possibly formed from a binary black hole merger, the observed spectrum of QNMs encodes information such as the angular momentum and mass of a black hole. Importantly, this encoding is independent of the initial conditions of perturbations, which allows for testing both general relativity and alternative theories of relativity \cite{Yunes2016}.
In particular,  recent studies have focused on mimicking  the geometry of a spinning  black hole in 1+2 dimensions using a draining bathtub vortex flow \cite{Visser1998,Richartz2015,Torres2017,Torres2020}.
 Subsequently, a theoretical extension to  binary Bose Einstein condensates (BECs) in 1+1 dimensions has been put forward as an analog model to simulate the behavior of the massive scalar field in the early Universe and black holes \cite{Fischer2004,Liberati2006,Visser2005,Syu2022,Syu2023,Syu2019}. 
 A further extension  to 1+2 dimensions provides a fantastic platform to study  quantum vortex instabilities in the background of  analogous rotating black holes \cite{Berti2023,Patrick2022}.\\

Black holes are typically not isolated but are at the centers of galaxies or surrounded by matter such as accretion disks and dark matter. Therefore, the spectra that we observe from gravitational waves should be modified by interactions with their surroundings \cite{Barausse2014,Cardoso2020,McGee2020}.
Recent studies in \cite{Cardoso2013,Cardoso2013b,Lingetti2022} have revealed  mechanisms that render Kerr black holes unstable due to the presence of matter in the vicinity of the black holes within the context of scalar-tensor theories, which alters the QNMs spectrum. One of the mechanisms involves positive mass squared due to the existence of so-called quasibound states, leading to superradiant instabilities  when the spin of a black hole exceeds a certain threshold.
The other comes from  negative mass squared, causing tachyonic instabilities that may further trigger the development of black-hole hairs, known as ``spontaneous scalarization''.

In this article, we consider binary BEC systems and introduce the BEC vortex. With  Rabi transition between two atomic hyperfine states, the
system can be represented by a coupled two-field model that involves both gapless excitations and gapped
excitations \cite{Fischer2004,Liberati2006,Visser2005,Syu2022,Syu2023}. 
The advent of  experimental studies on  the tunable binary BECs in \cite{Kim2020,Cominotti2022,Hamner2011,Hamner2013} makes it possible  to observe these two types of excitations.
 We  primarily focus  on the  gapped modes, which  are analogous to the massive scalar field.
 In the background of  zero and $\pi$ phase difference between two-component condensates, the induced mass term can be positive  and negative, respectively \cite{Zibold2010,Abbarchi2013,Tommasini2003}.
 Subsequently, the space-dependent Rabi coherent coupling  \cite{Hamner2013, Mitsunaga2000,Nicklas2011,Han2015} can generate a spatially dependent mass term, enabling the simulation of  superradiant and tachyonic instabilities  in the background of a draining  vortex flow with an analogous geometry to rotating black holes. 
 Our work opens up a new area of research to realize the effects of surrounding matter on black holes in a laboratory experiment using feasible spatially dependent Rabi coupling in binary BEC systems. 

\section{FORMALISM}\label{ABH2}
  We consider  tunable binary BECs with  identical atoms in two distinct internal hyperfine states.  Experimentally, one can tune the values of  scattering length  using Feshbach resonances, such as those associated with the two hyperfine states of $^{87}\text{Rb}$~\cite{Myatt1997,Hall1998,Papp2008,Tojo2010}. Additionally, we introduce the Rabi transition between these states with the strength determined by the Rabi frequency $\Omega$, where $\Omega\ge0$.
With the unit $\hbar=1$ throughout this paper, the coupled time-dependent equations of motion in $1+2$ dimensions are expressed by
\begin{align}\label{GP}
		 i\partial_t\hat{\Psi}_i=\left[-\frac{1}{2m_a}\nabla^2+V_i+g_{ii}\hat{\Psi}_i^\dagger\hat{\Psi}_i+g_{ij}\hat{\Psi}_j^\dagger\hat{\Psi}_j\right]\hat{\Psi}_i-\frac{\Omega}{2}\hat{\Psi}_j\,
\end{align}
with $i\neq j$ and $i,j=1,2$.  $m_a$ represents atomic mass and $V_i$ denotes the external potentials  on the hyperfine states $i$.
Furthermore, $g_{ii}, g_{ij}$ give the interaction strengths of atoms between the same hyperfine states and different hyperfine states, respectively. The coupling strengths  are related to the scattering lengths.
The condensate wave functions are given by the expectation value of the field operator $\langle\hat{\Psi}_i \rangle$ ,
$
	\langle \hat{\Psi}_i\rangle(\mathbf{ r},t)=\sqrt{\rho_{i}(\mathbf{ r},t)}\, e^{ i\theta_{i}(\mathbf{ r},t)-i \mu t}
$
 with the chemical potential $\mu$. The condensate flow velocities are given by $ \mathbf{v}_i={\grad} \theta_{i} /m_a$.

 The equations for $\rho_i$ and $\theta_i$ of the condensate wave functions can be found in \cite{Visser2005,Liberati2006,Syu2022}.
 The perturbations around the stationary wave functions are defined through
$
\hat{\Psi}_i=\langle\hat{\Psi}_i \rangle(1+\hat{\phi}_i) \, ,
$
{where 	the fluctuation fields decompose  in terms of the density and the phase  as
$
 		\hat{\phi}_i=\delta{\hat{n}_i}+i\delta\hat{\theta}_i=\frac{\delta\hat{\rho}_i}{2\rho_{i}}+i\delta\hat{\theta}_i\,.
$
According to \cite{Visser2005}, one can decouple the equations for the general spatially dependent condensate wave functions and the coupling strengths by choosing the time-independent background solutions $\rho_{1}=\rho_2=\rho$, $\theta_{1}=\theta_2=\theta$  or $\theta_{1}=\theta, \theta_2=\theta \pm \pi$, and
$g_{11}=g_{22}=g$ (see Appendix.~\ref{appendix1}) \cite{Zibold2010,Tommasini2003}. The  chosen scattering parameters of $g_{12}<g$ in the binary systems can result in a miscible state of the background condensates \cite{Kim2020,Hamner2011,Hamner2013}.

The decoupled equations are shown in \cite{Visser2005,Liberati2006,Syu2022}
in terms of the (gapless) density and (gapped) spin modes given by
$ \sqrt{2}\delta \hat n_{d/s}=\delta \hat{n}_1\pm\delta \hat{n}_2$ and
$ \sqrt{2}\delta \hat {\theta}_{d/s}=\delta \hat{\theta}_1 \pm\delta \hat{\theta}_2$.
Combining them yields a single equation for $\delta\hat\theta_s$
in the form of  the Klein-Gordon  equation for a massive scalar field,
\begin{align}
\frac{1}{\sqrt{-\mathbf{g}}}\partial_\mu\left(\sqrt{-\mathbf{g}}\,\mathbf{g}^{\mu\nu}\partial_\nu \delta\hat\theta_s\right)-m_\text{eff}^2\delta\hat\theta_s=0,
\label{KGE}
\end{align}
 where the spatial dependent sound speed  is given by $c^2=[(g-g_{12})\rho\pm\Omega]/m_a$, and  the effective mass by $m^2_\text{eff}=\pm 2m_a^2 c\,\Omega/\rho $, with the ``$+$'' sign corresponding to  the zero-phase difference and the ``$-$'' sign to the $\pi$-phase difference.
The acoustic metric $\mathbf{g}^{\mu\nu}$ depends on the background $\rho$ and $\theta$.
Following \cite{Patrick2022}, to mimic the geometry of a rotating black hole, we consider the draining vortex with the velocity flow
\begin{align}
	\mathbf{v}=\frac{1}{m_a}{\grad}\theta=\frac{-d \mathbf{e}_r+\ell \mathbf{e}_\phi}{m_a r}=v_r\mathbf{e}_r+v_\phi\mathbf{e}_\phi,
\end{align}
where the winding number $\ell$ is taken as an integer (hereafter $\ell=1$) for quantum vortices and $d>0$ for the draining flow.

In the Thomas-Fermi approximations,  the region of interest is  far from the core of the vortex with the radius $r \gg R_\text{TF}= \sqrt{\frac{d^2+l^2}{2m_a \rho_\infty(g+g_{12})}}$,
where  the density $\rho$ can be treated as a constant \cite{Patrick2022}, $\rho \simeq \rho_\infty$ given by
$
\rho_\infty=\frac{1}{g+g_{12}}\left(\mu-V_0 \pm\frac{\Omega}{2}\right)
$ for an uniform external potential $V_0$.
 The details can be seen in Appendix \ref{appendix1}. Using the coordinate transformations
$
	d\tilde{t}=dt-(c^2/v_r^2-1)^{-1}dr,
$
$
	d\tilde{\phi}=d\phi-\frac{v_rv_\phi}{r(c^2-v_r^2)}dr,
$
the metric becomes
\begin{align} ds^2&=\tilde g_{\mu\nu}^\text{vortex}d \tilde x^\mu  d \tilde x^\nu \nonumber\\
&\propto-\left(1-\frac{r_E^2}{r^2}\right)d\tilde{t}^2+\left(1-\frac{r^2_H}{r^2}\right)^{-1}dr^2\nonumber\\
&-\frac{2\ell}{m_a}\, d\tilde{\phi}d\tilde{t}+r^2d\tilde{\phi} ^2\, \label{metric}
\end{align}
with $\tilde x=(\tilde t, r, \tilde \phi)$. The acoustic horizon and the ergosphere are located at $r_\text{H}=d/m_ac$ and $r_\text{E}=\sqrt{d^2+\ell^2}/m_ac$, respectively.
Hereafter, we simplify the formula by defining  dimensionless variables as
$r/r_H\rightarrow r$, $c\,\tilde {t}/r_H\rightarrow  {t}$ and $r_H\Omega/c\rightarrow\Omega$.
Recently, the collective gapped excitation mode was experimentally observed in the binary BECs \cite{Cominotti2022}, prompting us to consider an experiment to simulate an analog black hole with environmental effects, which could be accessible by applying the space-dependent Rabi coupling \cite{Hamner2013,Mitsunaga2000,Nicklas2011,Han2015} in a two-dimensional BEC \cite{Fetter2001}.
We consider the following parameters:  the intraspecies scattering length $a_{11}=a_{22}=100a_0$, and the interspecies scattering length $a_{12}=90a_0$ , where $a_0$ is the Bohr radius. The atomic mass $m_a=1.44\times 10^{-25}\text{kg} $ and the asymptotic value of density $\rho_{\infty}$ can be prepared as much as $\sim 10^{21} \text{m}^{-3}$, 
giving the healing length and $R_\text{TF}$ such as $\xi \sim R_\text{TF} \sim 10^2 \text{nm}$. Considering $r_\text{H}\sim 10\xi$, from $v_r=c$, the value of $d$ is about  $d \gtrsim 1$. 
The corrections to the approximation of $\rho\simeq\rho_\infty$ are controlled by the small value of $R_\text{TF}/r_H$, which can be included beyond the constant density approximations in \cite{Oliveira2018,Demirkaya2020,Cardoso2022}.

Also, the measuring time to trace the evolution of the perturbed fields will be limited by the lifetime of the condensates, say $\sim 1\text{s}$, giving the
dimensionless time of order $ {t}\sim 10^{3}$.

Using the phase fluctuation field of the form
\begin{equation}
	\delta\theta_s({t},r,\tilde{\phi})=\frac{H_m (r, {t})}{\sqrt{r}}\exp\left(im\tilde{\phi}\right)
\end{equation}
in the  Klein-Gordon equation,  the radial equation in the tortoise coordinate
$
	r^\ast=r+\frac{1}{2}\log{\Big\vert\frac{r-1}{r+1}\Big\vert}
$ reads
{\begin{align}
 &\Bigg\{\frac{\partial^2}{\partial{r^\ast}^2}-\left(\frac{\partial}{\partial t}+i\frac{m\ell}{r^2}\right)^2
 -\left(\frac{r^2-1}{r^2}\right)\nonumber\\
 &\times\left[\frac{5}{4r^4}+\frac{1}{r^2}\left(m^2-\frac{1}{4}\right)+\mu_s^2(r)\right]\Bigg\}H_m(r,t)=0 \label{TDE}
\end{align}}
with $\mu^2_s(r)=\pm 2 d \Omega(r)$.
 Further writing $H_m (r,t)= H_{\omega,m}(r) e^{ -i\omega  {t}}$ turns the time-dependent equation into the time-independent Schrodinger-like equation
 \begin{align}
\left[\partial_{r^\ast}^2+V_\text{eff}(r)\right]H_{\omega,m}(r)=0,
\label{WEQ}
\end{align}
 where
\begin{align} V_\text{eff}(r)=&\left({\omega-\frac{m\ell}{r^2}}\right)^2-\left(\frac{r^2-1}{r^2}\right)\nonumber\\
	&\times\left[\frac{5}{4r^4}+\frac{1}{r^2}\left(m^2-\frac{1}{4}\right)+\mu^2_s(r)\right]\label{Veff2}
.
\end{align}
}
 Now the space-dependent mass term is introduced.
The mass term  due to the coupling to the surrounding matter of the black hole can be parametrized in terms of three parameters \cite{Cardoso2013,Cardoso2013b}.
One of them is  the magnitude of the mass reflecting the coupling strength between the scalar field and matter in the scalar-tensor gravity theories, which can be controlled by the Rabi coupling  $\Omega$.
The other two are related to the parametrization of the mass distribution. They are  the typical distance of the mass distribution $r_0$ from the black hole's horizon and the distribution width $1/\alpha$. The parametrization of the distance $r_0$ can be attributed to the radius of the so-called  innermost stable circular orbit of the particle moving around a massive object such as a black hole. The radius of the innermost circular motion depends on the mass and angular momentum (spin) of the massive object. The width of the mass distribution $1/\alpha$ is also an important parameter characterizing the surrounding matter of the black holes.

Compared to the counterpart of the effective potential defined from the Klein-Gordon equations in the scalar-tensor gravity theories, the existence of the ergosphere in the background metric is essential to account for the pure ingoing wave boundary condition at the horizon for $\omega < ml$. The presence of the mass term provides a potential barrier, leading to the possibility of  quasibound states. Also, as $r\rightarrow \infty$,  $V_\text{eff}(r) \rightarrow \omega^2$, where the pure outgoing wave solution can be achieved. 

\begin{figure}[t]
	\includegraphics[width=1\linewidth]{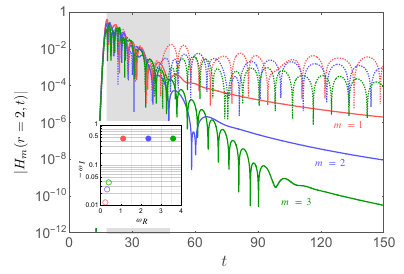}
	\caption{The comparison of the absolute value of the time-dependent profile between the cases of $\Omega_0=0.0$ (solid line) and $\Omega_0=0.4$ (dashed line) with $m=1$ (red), $m=2$ (blue), and $m=3$ while the  distance is set to $r_0=15$.
The shaded time regime indicates  the echo time with $\Delta t\simeq 2r_0$.
The inset figure shows the fundamental QNM frequency (filled circle $\rightarrow$ solid line, open circle $\rightarrow$ dashed line) obtained from frequency domain. }
	\label{fig_time_evo}
\end{figure}
\begin{figure}[t]
	\includegraphics[width=1\linewidth]{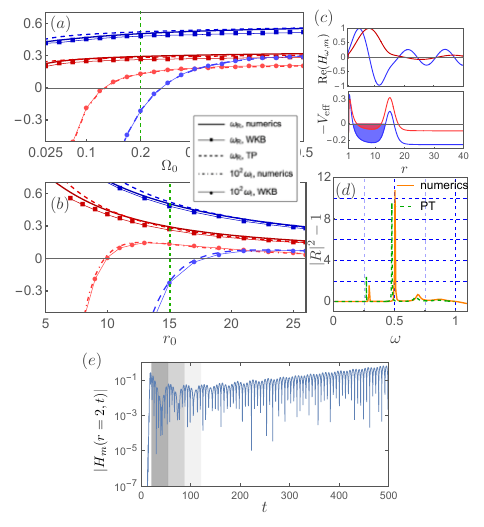}
	\caption{In the case of  positive mass squared, (a) shows the frequency for the mode $n=0$ (red) and $n=1$ (blue) as a function of $\Omega_0$ with fixed $r_0=15$, and (b) shows the corresponding frequency  as a function of $r_0$ with $\Omega_0=0.2$.  (c) depicts the effective potential and the normalized radial profile  for the first two modes, and  (d)
	presents a series of peaks at the resonant frequencies in the reflection coefficient spectrum.
	 (d) reveals the growing waveform in time domain and the echo time duration. (c)--(e) share the same values of $\Omega_0$ and $r_0$ marked in the green line in (a) and (b). Note that $m=1$, and $\alpha=1/2$ are set in all figures.}
	\label{fig_spectrum}
\end{figure}
Given the boundary conditions of the pure ingoing wave $e^{-i(\omega-m\ell)r^\ast}$ at the horizon and the pure outgoing wave $e^{i\omega r^\ast}$ at infinity, the eigenfrequency of the time-independent Schrodinger-like equation can be found with  the complex values  $\omega=\omega_R+i\omega_I$.
However, there are two  types of instabilities of interest with  very different dynamics on the timescale $\tau\propto1/\vert\omega_I\vert$: superradiant instabilities \cite{Patrick2018} are driven by an unstable but  long-lived mode where $\omega_R\gg \omega_I$
, while tachyonic instabilities are featured with an almost pure imaginary frequency $\omega_I\gg \omega_R$ leading to a short-lived mode.
In two-component BECs,  the effective mass squared term can be parametrized as an analytically treatable form
\begin{align}
	\mu^2_s (r)=\pm2d\Omega(r)=\pm 2d\Omega_0\,\text{sech}^2[\alpha(r-r_0)],\label{mass}
\end{align}
 mimicking a mass shell surrounding the black hole \cite{Cardoso2013,Cardoso2013b,Cheung2022,Courty2023}. The same instabilities are expected to be  observed for
 a more general parametrization of the mass distribution, controlled by its typical distance from the black hole horizon and the width of the distribution. 
 
 However, there are other types of instabilities called ergoregion instabilities in \cite{Giacomelli2020,Giacomelli2021}, which are not caused by a trap in the effective potential.
 
  In the case of the spatially dependent mass term \eqref{mass},
we calculate the eigenfrequency numerically using the shooting method \cite{Cardoso2005}, where the solutions are obtained by integrating the equation from the horizon onward and  from the far distance $r$ backward with appropriate boundary conditions where  they continuously vary over $r$.
We also find the eigenfrequency  with the  generalized P\"{o}schl-Teller (PT) semianalytical methods \cite{Macedo2018, Torres2022, Casals2009}.  Additionally, the modified continued fraction method is adopted to obtain QNM's frequencies shown in Appendix \ref{appendix2} \cite{Leaver1990}.
The  analytically approximated formulas for the eigenfrequency are given below. The instabilities are shown in the time domain by solving the time-dependent equation (\ref{TDE}).
For technical details, refer to Appendix \ref{appendix3}.

\section{Quasinormal modes and superradiant instabilities.}
In the case of the quasinormal modes, their spectrum   is strongly modified by the presence of the spatially dependent mass shell, which introduces an additional  barrier to the rotational barrier given by the azimuthal angular momentum $m$.
The frequencies of quasinormal modes for the real and imaginary parts are found in Fig.~\ref{fig_time_evo}, which are  smaller than their counterparts without the mass shell  \cite{Berti2004, Cardoso2004}.
 The corresponding time domain signals are obtained  by solving the time-dependent equation (\ref{TDE}) for an initial Gaussian function with reference to \cite{Berti2022}. According to \cite{Berti2022}, small perturbations can lead to a relatively large shift of the eigenfrequency from the unperturbed one, but not  to a significant change in the time domain (see an example in Appendix \ref{appendix2}). Here, we present the evolution of the radial profiles,  which show a significant change due to the nonzero Rabi coupling strength.
The echo time (estimated from the shaded time domain around $\sim 2 r_0$), which  is the time for the radial profile to travel towards the potential barrier at $r=r_0$ and return to the starting position and the ringdown of the profile at late times are displayed in Fig \ref{fig_time_evo}.
More interestingly, when  $\Omega_0$ or $r_0$ reaches its respective critical value, the  positive mass shell could further induce  superradiant instabilities,  where a  local bound state is shown for the potential ($- V_\text{eff}$) in Fig.~\ref{fig_spectrum}. This is similar to superradiant instabilities induced by the accretion disk around Kerr black hole  in scalar-tensor gravity theories \cite{Cardoso2013,Cardoso2013b,Lingetti2022}.
Let us  estimate the resonance spectrum via a discrete quantization condition in the Wentzel-Kramers-Brillouin (WKB) analysis given by
\begin{align}
	\int_{r^{\ast}_-}^{r^{\ast}_+} dr^\ast \sqrt{V_\text{eff}(r,r_0,\Omega_0,\alpha)}=\left(
	n+\frac{1}{2}\right)\pi,\quad n=0,1,2\cdots, \label{WKB}
\end{align}
where $r_{\pm}^\ast$ are the classical turning points in the effective potential $-V_\text{eff}$.
Considering $\omega_R\gg\vert\omega_I\vert$, the condition \eqref{WKB} can be approximated as
\begin{align}
	-\frac{1}{2} \pi  \sqrt{2 m\ell \omega_R +m^2}+\zeta  \omega_R +\frac{m^2}{2 \zeta  \omega_R }=(n+\frac{1}{2})\pi
\end{align}
as $r_0\gg r_H$ with $\zeta=r_0-\alpha^{-1}{\ln \left(\frac{2 \sqrt{2d\Omega_0}}{\omega_R }\right)}$ and $\Omega_0>0$,
from which $\omega \simeq \omega_R$ can be determined.
In the limit of a thin width of the mass shell where $\alpha\rightarrow \infty$,  $\zeta\rightarrow r_0$, the real part of the frequency becomes $\omega_R\sim(\frac{m}{2}+n+\frac{1}{2})\pi/r_0$ reducing to the known result in \cite{Cardoso2004b}.
{After plugging in the resulting frequency $\omega\simeq \omega_R$ in $V_\text{eff}$,
 increasing $r_0$  shifts the barrier away from the rotational barrier so that  the shaded area
in Fig.~\ref{fig_spectrum}  decreases, giving a smaller value of $\omega_R$.}

 Nevertheless, it is noted  that, for example, for $n=0$ (the fundamental mode) and $n=1$ (the overtone mode), if $r_0$ is above  a certain value shown in Fig.~\ref{fig_spectrum}, the imaginary part of the eigenfrequency $\omega_I$ becomes positive, indicating  superradiant instabilities due to the existence of  the so-called quasibound state. Then, as $r_0$ falls below the critical value, the negativeness  of $\omega_I$ reveals the quasinormal mode.
{On the contrary, as  $\Omega_0$ increases, the shaded area increases, resulting in a larger value of $\omega_R$. Similarly, there is a critical value of $\Omega_0$, above (below) which the mode suffers (becomes) superradiant instabilities (quasinormal modes).}
In this setup, sending a monochromatic wave   into the vortex will be scattered  even if the mass shell shields the vortex. If its frequency is $\omega<m\ell$, then it will cause  superradiant amplification, and if the frequency is also close to the spectrum frequency $\omega\sim \omega_{R}$, significant resonance amplification will occur, as shown in Fig.~\ref{fig_spectrum}.  Either a stable quasinormal mode or an unstable quasibound state is partially trapped within the mass shell and partially propagate towards  spatial infinity,  resembling  a new class of  modes in the astrophysics \cite{Cardoso2013b}. The instabilities associated with the quasibound states are also evident from the growth of the radial profile with time in Fig.~\ref{fig_spectrum}. While the function $H_m (r=2,t)$ grows, the echo time of two neighboring peaks is about $2 r_0$. Within the lifetime of the condensates, instabilities will destabilize  the background solutions, where  backreaction effects must be systematically taken into account.
\begin{figure}[t]
	\includegraphics[width=1\linewidth]{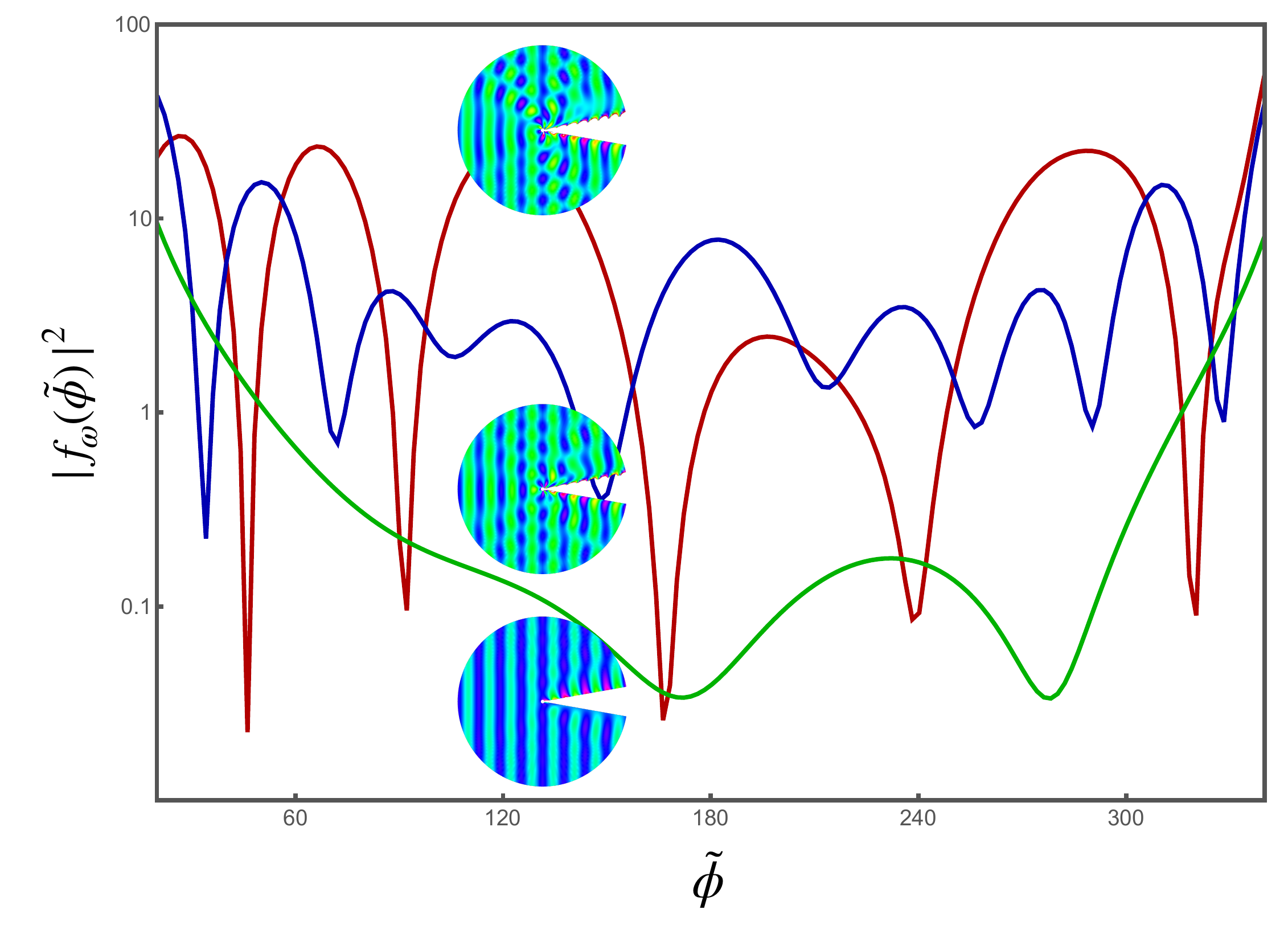}
	\caption{The absolute square of the scattering amplitude as a function of $\tilde\phi$ for frequency $\omega=0.5$, with $\Omega_0=0.2$ (red)(quasibound state frequency shown in Fig.~\ref{fig_spectrum}), and $\Omega_0=0.1$ (blue), $\Omega_0=0.0$ (green). The interference pattern given by the real part of $ e^{i\omega x}+f_\omega(\tilde\phi)\frac{e^{i\omega r^\ast}}{\sqrt{r}}$ is shown in inset figures.}
	\label{fig_scattering}
\end{figure}

Another method to justify the existence of the quasibound states is through scattering experiment where time dependent measurements might face challenges due to its long time scale $\tau \propto1/\vert\omega_I\vert$. Therefore, we investigate scattering to see the effect of mass shell around the black hole on the imprint carried by the scattered wave. Scattering problems for the scalar field  in the draining bathtub are intensively  studied \cite{Oliveira2010,Dolan2012,Dolan2013b}. The differential cross section is given by
	\begin{align}
		\frac{d\sigma}{d\tilde\phi}=|f_\omega(\tilde\phi)|^2,
	\end{align}
where the scattering amplitude $f_\omega$ obeys the solution of the Klein-Gordon equation in (\ref{KGE}) with the metric (\ref{metric})
\begin{align}
	\delta\theta_s(t,r,\tilde\phi)=e^{-i\omega t}\left[e^{i\omega x}+f_\omega(\tilde\phi)\frac{e^{i\omega r^\ast}}{\sqrt{r}}\right].
\end{align}
We present the results of the scattering cross section $d\sigma/d\tilde\phi$ obtained from the partial wave summation in Fig.~\ref{fig_scattering}. We find that
the large angular scattering occurs for the  scalar field with nonzero mass term compared to the zero mass term cases \cite{Dolan2012, Dolan2013}. In the case of the quasibound state, the absolute square of the scattered amplitude
 due to resonance amplification becomes even larger than nonresonance cases. The corresponding the interference patterns that might be observed in experiments
 given by the real part of $e^{i\omega x}+f_\omega(\tilde\phi)\frac{e^{i\omega r^\ast}}{\sqrt{r}}$ are also displayed in the inset figures.

\begin{figure}[t]
	\includegraphics[width=1\linewidth]{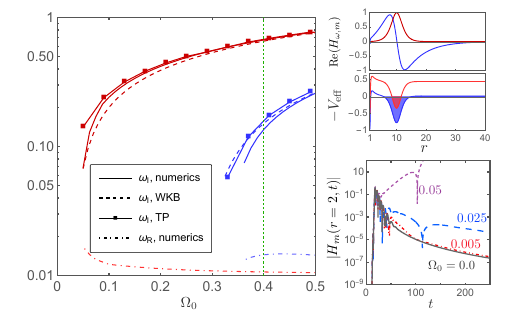}
	\vspace*{-7mm}
	\caption{
		The left panel shows the comparison among numerics, WKB  and PT methods for the imaginary part of frequency as a function of $\Omega_0$.
		The right top panels depict the effective potential and the normalized radial profile for modes $n=0,1$. The right bottom panel reveals the time-dependent profile with varying $\Omega_0$. The parameters are fixed as $m=1,\,r_0=10$ and $\alpha=1/2$.}
	\label{fig_tachyon}
\end{figure}
\section{Tachyonic instabilities} A negative mass squared could trigger  tachyonic instabilities in scalar-tensor theories, which further leads to spontaneous scalarization.
Tachyonic instabilities arise from a completely different mechanism than that of superradiant instabilities.
It is expected that tachyonic instabilities lead to a large imaginary part with a short lifetime, namely $ \omega_I \gg \omega_R$. We present numerics, PT, and WKB analytical methods to confirm the existence  of the tachyonic instability spectrum  in Fig.~\ref{fig_tachyon}.  A fundamental mode ($n=0$) has the largest frequency $ \omega\simeq\omega_I$  for $\omega_I >0$ compared with the overtones. A larger coupling $ \Omega_0 $ and a wider shell $\alpha^{-1}$ lead to stronger tachyonic instabilities. The radial profiles $H_{\omega,m}(r)$ are also drawn for $n=0$, the fundamental mode, and $n=1$, the overtone mode, which lie within the respective classical allowed regimes.
Let us assume that \eqref{WKB} can still be applied to give an analytical formula for the eigenfrequency with $\omega \simeq i\omega_I$. Substituting $\omega\simeq i\omega_I$ into $V_\text{eff}$  shows the potential profile in Fig.~\ref{fig_tachyon}. The shaded area for the classical allowed region obeys (\ref{WKB})  for an integer $n$ given by
\begin{align}
	\omega_I\sim\frac{1}{3}\left[\sqrt{2d\Omega_0}-\alpha\left(n+\frac{1}{2}\right)\right]\pi . \label{omega_I_WKB}
\end{align}
The values of $\omega_I$ are  in good agreement with the other methods in Fig.~\ref{fig_tachyon}. The change in $r_0$  may shift the potential profile horizontally  but keep the  shaded area almost unchanged, giving little effect to the eigenfrequencies.
 The value of $\Omega_0$   shifts the potential profile vertically, and
 from  the analytical formulas in (\ref{omega_I_WKB}), the condition  $\sqrt{2d\Omega_0}> \alpha ( n+1/2)$ triggers tachyonic instabilities, resulting in $\omega_I >0$. Otherwise, the resulting $\omega_I < 0$ indicates the quasinormal modes.
 The corresponding evolution of the radial profile from \eqref{TDE} is also displayed in the cases of the quasinormal modes for relatively small $\Omega_0$ and the tachyonic instabilities for relatively larger $\Omega_0$.
\section{Conclusion}
We conclude this article by summarizing the parameter constraints for a successful analog gravity model. The zero-phase  difference between the two types of  condensates give perturbed fields with positive mass squared, and the $\pi$-phase difference can give negative mass squared where the choice of one of two modes allows for experimental feasibility \cite{Zibold2010,Abbarchi2013}.
Positive sound speed squared is required, where $g_{12}/g < 1\pm \Omega_0/\rho g$ (with a dimension-restored variable $\Omega_0$),  leading to a
further constraint on $\Omega_0 <\rho (g-g_{12})$ in the $\pi$-phase difference modes \cite{Bernier2014,Recati2022}.
 The experimentally challenging but achievable spatially dependent Rabi coupling strength has been discussed in \cite{Nicklas2011}.

\section{acknowledgments}
	This work was supported in part by the National Science and Technology Council (NSTC) of Taiwan, Republic of China.
\appendix

\section{BACKGROUND SOLUTIONS AND THOMAS-FERMI APPROXIMATIONS }\label{appendix1}
We show the detailed derivations of the background solutions for  the density and phase of two-component BECs.
Plugging in the parametrization of the condensate wave functions of $\langle \hat{\Psi}_i\rangle(\mathbf{r},t)=\sqrt{\rho_{i}(\mathbf{r},t)}\, e^{ i\theta_{i}(\mathbf{r},t)-i \mu_i t}$ to  \eqref{GP}, the density and phase satisfy the equations given, respectively by
\begin{widetext}
\begin{subequations}
\begin{align}
	&i\partial_t\left(\sqrt{\rho_1}e^{i(\theta_1-\mu_1t)}\right)=\left(
	-\frac{1}{2m_a}\nabla^2+V_{\text{ext},1}+g_{11}\rho_1+g_{12}\rho_2
	\right)\sqrt{\rho_1}e^{i(\theta_1-\mu_1t)}-\frac{\Omega}{2}\sqrt{\rho_2}e^{i(\theta_2-\mu_2t)},\\
	&i\partial_t\left(\sqrt{\rho_2}e^{i(\theta_2-\mu_2t)}\right)=\left(
	-\frac{1}{2m_a}\nabla^2+V_{\text{ext},2}+g_{22}\rho_2+g_{12}\rho_1
	\right)\sqrt{\rho_2}e^{i(\theta_2-\mu_2t)}-\frac{\Omega}{2}\sqrt{\rho_1}e^{i(\theta_1-\mu_1t)}.
\end{align}
\end{subequations}
In the paper, we focus on the miscible regime and choose the parameters in  \eqref{GP} as
\begin{align}
	&g_{11}=g_{22}\equiv g,\quad g_{12}<g,\\
	&\mu_1=\mu_2\equiv\mu,\quad V_1=V_2\equiv V_0.
\end{align}
Using the fact that
\begin{align}
	i\partial_t\left(\sqrt{\rho_1}e^{i(\theta_1-\mu_1t)}\right)=\left(
	-\frac{1}{2m_a}\nabla^2+V_0+g\rho_1+g_{12}\rho_2
	\right)\sqrt{\rho_1}e^{i(\theta_1-\mu_1t)}-\frac{\Omega}{2}\sqrt{\rho_2}e^{i(\theta_2-\mu_2t)}, \label{GP3}
\end{align}
the left-hand side of \eqref{GP3} gives
\begin{align}
	i\left[\partial_t\sqrt{\rho_1}+i\sqrt{\rho_1}(\partial_t\theta_1-\mu )\right]e^{i(\theta_1-\mu t)}
\end{align}
and the right-hand side of \eqref{GP3} becomes
\begin{align}
	&\Big\{-\frac{1}{2m_a\sqrt{\rho_1}}\Big[\nabla^2\sqrt{\rho_1}+i\nabla\sqrt{\rho_1}\cdot\nabla\theta_1+i\sqrt{\rho_1}\nabla^2\theta_1
	+i\nabla\theta_1\cdot\left(\nabla\sqrt{\rho_1}+i\sqrt{\rho_1}\nabla\theta_1\right)\Big]\nonumber\\
	&\quad\qquad V_{0}+g_{11}\rho_1+g_{12}\rho_2\Big\}\sqrt{\rho_1}e^{i(\theta_1-\mu t)}-\frac{1}{2}\sqrt{\rho_2}e^{i(\theta_2-\mu t)}.
\end{align}
Therefore, its imaginary part of \eqref{GP3} leads to
\begin{subequations}\label{comp1}
	\begin{align}
		&\partial_t\sqrt{\rho_1}=\frac{\dot{\rho}_1}{2\sqrt{\rho_1}}\nonumber\\
		&=-\frac{1}{2m_a\sqrt{\rho_1}}\Big(\nabla\sqrt{\rho_1}\cdot\nabla\theta_1+\sqrt{\rho_1}\nabla^2\theta_1
		+\nabla\theta_1\cdot\nabla\sqrt{\rho_1}\Big)\sqrt{\rho_1}-\frac{\Omega}{2}\sqrt{\rho_2}\sin{(\theta_2-\theta_1)}\nonumber\\
		&\Rightarrow\nonumber\\
		&\dot{\rho}_1=
		-\frac{1}{m_a}\Big(2\nabla\sqrt{\rho_1}\cdot\nabla\theta_1+\sqrt{\rho_1}\nabla^2\theta_1
		\Big)\sqrt{\rho_1}-{\Omega}\sqrt{\rho_1\rho_2}\sin{(\theta_2-\theta_1)}\, ,
	\end{align}
	whereas the real part gives
	\begin{align}
		&\sqrt{\rho_1}(\dot{\theta}_1-\mu)=\frac{1}{2m_a}\left[\nabla^2\sqrt{\rho_1}-\sqrt{\rho_1}(\nabla\theta_1)^2\right]-\left(V_0+g\rho_1+g_{12}\rho_2\right)\sqrt{\rho_1}\nonumber\\
		&\qquad\qquad\qquad+\frac{\Omega}{2}\sqrt{\rho_2}\cos{(\theta_2-\theta_1)}\nonumber\\
		&\Rightarrow\nonumber\\
		&\dot{\theta}_1=\frac{1}{2m_a\sqrt{\rho_1}}\left[\nabla^2\sqrt{\rho_1}-\sqrt{\rho_1}(\nabla\theta_1)^2\right]-\left(V_0+g\rho_1+g_{12}\rho_2-\mu\right)+\frac{\Omega}{2}\sqrt{\frac{\rho_2}{\rho_1}}\cos{(\theta_2-\theta_1)}.
	\end{align}
\end{subequations}
Similarly, the counterpart coupled equations for the  hyperfine state $2$ have the form
\begin{subequations}\label{comp2}
	\begin{align}
		&\dot{\rho}_2=
		-\frac{1}{m_a}\Big(2\nabla\sqrt{\rho_2}\cdot\nabla\theta_2+\sqrt{\rho_2}\nabla^2\theta_2
		\Big)\sqrt{\rho_2}-{\Omega}\sqrt{\rho_1\rho_2}\sin{(\theta_1-\theta_2)},\\
		&\dot{\theta}_2=\frac{1}{2m\sqrt{\rho_2}}\left[\nabla^2\sqrt{\rho_2}-\sqrt{\rho_2}(\nabla\theta_2)^2\right]-\left(V_0+g\rho_2+g_{12}\rho_1-\mu\right)+\frac{\Omega}{2}\sqrt{\frac{\rho_1}{\rho_2}}\cos{(\theta_1-\theta_2)}.
	\end{align}
\end{subequations}
The stationary solutions satisfy $\dot{\rho}=\dot{\theta}=0$ where the dot means the derivative with respect to time $t$. One would therefore immediately obtain the solutions, which are
\begin{align}
	\rho_1=\rho_2=\rho,\qquad\theta_1=\theta \, ,\theta_2=\theta  \,; \quad\quad \theta_1=\theta \, ,\theta_2=\theta\pm\pi \, 
	\label{cond1}
\end{align}
for zero-phase difference  and $\pi$-phase difference.
Imposing \eqref{cond1} into \eqref{comp1} and \eqref{comp2} gives
\begin{subequations}\label{EOM_simplified}
	\begin{align}
		&-\frac{1}{m_a}\Big(2\nabla\sqrt{\rho}\cdot\nabla\theta+\sqrt{\rho}\nabla^2\theta
		\Big)\sqrt{\rho}=0\nonumber\\
		&\qquad\qquad\Rightarrow-\nabla\cdot\left(\rho\frac{\nabla\theta}{m_a}\right)=0\Rightarrow\nabla\cdot(\rho \vec{v})=0,\label{EOM_simplified1}\\
		&\frac{1}{2m_a\sqrt{\rho}}\left[\nabla^2\sqrt{\rho}-\sqrt{\rho}(\nabla\theta)^2\right]-\left[V_0+(g+g_{12})\rho-\mu\right]\pm\frac{\Omega}{2}\nonumber\\
		&\qquad\qquad\Rightarrow \frac{1}{2m_a\sqrt{\rho}}\nabla^2\sqrt{\rho}-\frac{m_a}{2}\mathbf{v}^2-V_0-(g+g_{12})\rho+\mu\pm\frac{\Omega}{2}=0,\label{EOM_simplified2}
	\end{align}
\end{subequations}
\end{widetext}
where we have used the definition of the condensate flow velocity $\mathbf{v}=\grad \theta/m_a$. The first term in \eqref{EOM_simplified2} is  the quantum pressure to be ignored in the hydrodynamical regime, and the plus and minus signs of the last term correspond to the solutions of zero and $\pi$ phase differences, respectively.

To solve the stationary solution, we  consider the flow velocity
\begin{align}
	\mathbf{v}=-\frac{d}{m_a r}\mathbf{e}_r+\frac{\ell}{m_a r}\mathbf{e}_\phi,
\end{align} then in the hydrodynamical limit  \eqref{EOM_simplified2} becomes
\begin{align}
	&-\left(\frac{d^2+\ell^2}{2m_a}\right)\frac{1}{r^2}-V_0-(g+g_{12})\rho+\mu\pm\frac{\Omega}{2}=0.
\end{align}
Considering uniform potential $V_0$, one can find the density as
\begin{align}
	\rho(r)=\rho_\infty(1-\frac{R_\text{TF}^2}{r^2})\label{density}
\end{align}
with the asymptotic density
\begin{align}
	\rho_\infty=\frac{1}{g+g_{12}}\left(\mu-V_0\pm\frac{\Omega}{2}\right),
\end{align} and the radius of the vortex core
\begin{align}
	R_\text{TF}^2= \frac{d^2+\ell^2}{2m_a \rho_\infty(g+g_{12})}.\label{vortex_radius}
\end{align}

In the text, we choose the experimental parameters so that the Thomas-Fermi radius $R_\text{TF}$ is far smaller than the
horizon radius $r_H$, namely $r_H\gg R_\text{TF}$, where the density can be safely treated as a constant $\rho(r>r_H)\simeq\rho_\infty$ \cite{Patrick2022}. Regarding to the density beyond constant approximation, the superradiant amplification could be modified \cite{Oliveira2018,Demirkaya2020,Cardoso2022}.

\section{P\"{O}SCHL-TELLER METHOD AND CONTINUED FRACTION METHOD}\label{appendix2}
\begin{figure}[tb]
	\includegraphics[width=\linewidth]{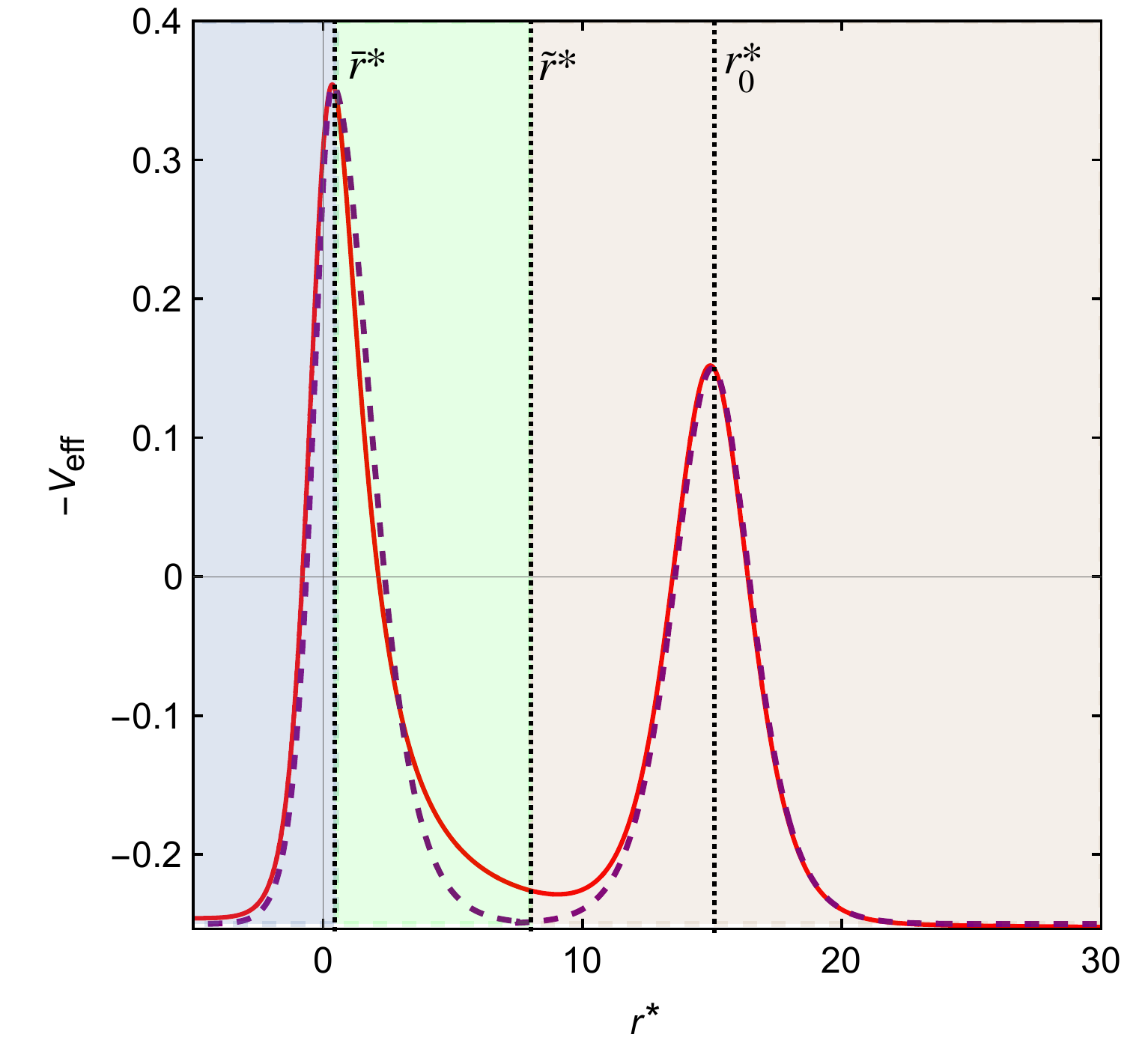}
	\caption{Comparison between the typical effective potential $V_\text{eff}$ (red) defined in the text and the generalized P\"{o}schl-Teller potential (purple dashed). We show three regions with different background colors where interface boundary indicates the matching locations.}
	\label{fig_TPmethod}
\end{figure}

A relatively small mass shell near the black hole could significantly affect the distributions of quasinormal mode spectra and even leads to instabilities. To examine this discovery, we may seek the solution satisfying a pure incoming wave at the horizon and a pure outgoing wave at infinity expressed as  a linear combination of
two functions $h_H$ and $h_\infty$, given by
\begin{align}
	h_H=\bigg\{\begin{array}{ll}
		e^{-i(\omega-m\ell)r^\ast},\qquad&r^\ast\rightarrow -\infty\\
		A_\infty^-e^{-i\omega r^\ast}+A_\infty^+e^{i\omega r^\ast},&r^\ast\rightarrow \infty
	\end{array}\label{hH},
\end{align}
and
\begin{align}
	h_\infty=\bigg\{\begin{array}{ll}
		A_H^-e^{-i(\omega-m\ell)r^\ast}+A_H^+e^{i(\omega-m\ell)r^\ast},\qquad&r^\ast\rightarrow -\infty\\
		e^{i\omega r^\ast},&r^\ast\rightarrow \infty
	\end{array}
	\label{hinfty},
\end{align}
respectively. The spectrum then can be obtained from the condition $A_H^+/A_H^-=0$ \cite{Macedo2018, Torres2022}.
According to the effective mass term we chose in \eqref{mass}, it is suitable to adopt the generalized P{\"o}schl-Teller potential by splitting a whole space into three regions through the following parametrization \cite{Torres2022}
\begin{align} V_\text{PT}(r^\ast)=&V_3(r^\ast)\Theta(\bar{r}^\ast-r^\ast)\nonumber\\&+V_2(r^\ast)[\Theta(r^\ast-\bar{r}^\ast)-\Theta(r^\ast-\tilde{r}^\ast)]\nonumber\\&+V_1(r^\ast)\Theta(r^\ast-\tilde{r}^\ast)\,,
	\label{PT}
\end{align}
where
\begin{align}
	&V_{i}(r^\ast)=a_{i}+b_{i}\text{sech}^2{[(r^\ast-{r}_i^\ast)/s_{i}]}\,
\end{align}
with $r_1^\ast=r_0^\ast, r_2^\ast=\tilde{r}^\ast, r_3^\ast=\bar{r}^\ast$.   $\bar{r}^\ast$ is the position of the top of the rotational barrier, and $\tilde{r}^\ast$ is selected to be within the interval $\{\bar{r}^\ast,r_0^\ast\}$ where  $r_0^\ast$ is the position of another barrier induced by the mass term in Fig.~\ref{fig_TPmethod}.
The parameters can be determined by comparing with the effective potential $V_\text{eff}$  with the mass term in \eqref{mass} in the text.
The coefficients $a_1$ and $b_1$ can also be determined by $ r^\ast \rightarrow \infty$ and the mass shell formula giving $a_1=\omega^2$ and $b_1=-2 \Omega_0$. Likewise, the coefficients $a_1$ and $b_1$ are obtained from the effective potential near the horizon as well as at the top of the rotational barrier giving $a_3=(\omega-ml)^2$ and $b_{3}=V_\text{eff}(\bar{r}^\ast)-a_{3}$.  Finally, considering the barrier given by the mass shell is far from the rotational barrier, $a_2=\omega^2$ and the coefficient $b_2$ can be related to the potential value at the rotational barrier by $b_{2}=V_\text{eff}(\bar{r}^\ast)-a_{2}$. The width $s_3$ is the mass shell width $1/\alpha$.
The widths $s_{1,2}$ are chosen to make $V_\text{PT}$ is differentiable at the top of barrier $\bar{r}^\ast$ \cite{Torres2022}.
In each regions, the general solution to the Schrodinger-like equation
\begin{align}
	\partial_{r^\ast}^2H (  r^\ast)+V_i H (r^\ast)=0
\end{align}
is given by
the linear combination of two independent solutions  \cite{Casals2009}
\begin{align}
	H_{ i}(r^\ast)=A_{i}h_{i}^\text{in}(r^\ast)+B_{i}h_{i}^\text{up}(r^\ast), \label{Hi_eq}
\end{align}
where $h^\text{(in/up)}_{i}(r^\ast)=\Gamma(1-\mu_{i})P_{\beta_{i}}^{\mu_{i}}(\mp\tanh(r^\ast-{r_i}^\ast)/s_{i})$ obeys the asymptotic behaviors
$
h_i^{(\text{in})}\sim e^{-\mu_i r^\ast/s_i}
$ as $r^\ast\rightarrow -\infty$,  and $
h_i^{(\text{up})}\sim e^{\mu_i r^\ast/s_i}
$ as $r^\ast\rightarrow \infty$.
The parameters $\mu_i$ and $\beta_i$ are given by
\begin{align}
	&\mu_{i}={is_{i}\sqrt{a_{i}}},\nonumber\\
	&\beta_{i}=-\frac{1}{2}+\frac{\sqrt{1+4s_{i}^2b_{i}}}{2}.
\end{align}
Given the condition of (\ref{hinfty}) at $r^\ast \rightarrow \infty$, it requires $A_1=0,B_1=1$.
We then match the functions $h_1^{\text{(up)}}$, $h_2^{\text{(in)}}$, $h_2^{\text{(up)}}$, $h_3^{\text{(in)}}$ and $h_3^{\text{(out)}}$ at two matching points: one is the top of the rotational barrier $\bar{r}^\ast$ and the other is $\tilde{r}^\ast$, say at $\tilde{r}^\ast=({r}_0^\ast+\tilde{r}^\ast)/2$ given by
\begin{align}
	&h_1^\text{(up)}\big\vert_{\tilde{r}^\ast}={A}_{2}h_2^\text{(in)}\big\vert_{\tilde{r}^\ast}+{B}_{2}h_2^\text{(up)}\big\vert_{\tilde{r}^\ast},\\
	&\partial_{r^\ast} h_1^\text{(up)}\big\vert_{\tilde{r}^\ast}={A}_{2}\partial_{{r}^\ast} h_2^\text{(in)}\vert_{\tilde{r}^\ast}+{B}_{2}\partial_{{r}^\ast} h_2^\text{(up)}\big\vert_{\tilde{r}^\ast},\\
	&{A}_{3}h_3^\text{(in)}+{B}_{3}h_3^\text{(up)}\big\vert_{\bar r^\ast}=\mathcal{A}_{2}h_2^\text{(in)}\big\vert_{\bar r^\ast}+{B}_{2}h_2^\text{(up)}\big\vert_{\bar r^\ast},\\
	&{A}_{3}\partial_{r^\ast}h_3^\text{(in)}\big\vert_{\bar r^\ast}+{B}_{3}\partial_{r^\ast}h_3^\text{(up)}\big\vert_{\bar r^\ast}\\\nonumber
	&\qquad\qquad\qquad={A}_{2}\partial_{r^\ast}h_2^\text{(in)}\big\vert_{\bar r^\ast}+{B}_{2}\partial_{r^\ast}h_2^\text{(up)}\big\vert_{\bar r^\ast}.
\end{align}
Now the coefficients $A_3$ and $B_3$ can be determined from above matching conditions.
To read out $A_H^+$ and $A_H^-$ for the asymptotical behavior of the solution at $r^\ast \rightarrow -\infty$ in (\ref{hinfty}) from the solution at $i=3$ in (\ref{Hi_eq}), one more step is to rewrite $h_3^\text{(up)}$ into a combination of the following plane wave forms by employing the identity of the hypergeometric functions
\begin{widetext}
	\begin{table*}[t]
		\label{tb_QNM}
		\centering
		\caption{The comparison of the fundamental mode frequency with the shooting method (SM) and  continued fraction method (CFM). We fixed $m=1$, $\alpha=1/2$, $r_0=15$.}
		
		\begin{tabular}{ccccc}
			\hline
			\hline
			\\[-6pt]
			\multicolumn{1}{c}{} & \multicolumn{2}{c}{\textbf{$\Omega_0=0.1$}} & \multicolumn{2}{c}{\textbf{$\Omega_0=0.2$}} \\
			\cmidrule(rl){2-3} \cmidrule(rl){4-5}
			\textbf{$m$} & {SM} & {CFM} & {SM} & {CFM} \\
			\midrule
			-2 & {$0.3027-0.0055 i$} & {$\,\,\,\,0.3029-0.0055i$}  & {$\,\,0.3231 - 0.0031 i$}  &{$\,\,0.3231-0.0031$}  \\
			-1 & {$0.2238-0.0136i$} & {$\,\,0.2238-0.0137i$}  & {$\,\,0.2392 - 0.0152 i$}  & {$\,\,0.2392-0.0152i$}  \\
			1 & {$0.2694-0.0019i$} & {$\,\,0.2696-0.0020i$}  & {$\,\,0.2914 + 0.0013 i$}  & {$\,\,0.2915 + 0.0013i$}  \\
			2 & {$0.3602-0.0076i$} & {$\,\,0.3605-0.0075i$}  & {$\,\,0.3906 - 0.0010i$}  & {$\,\,0.3907 - 0.0011 i$} \\
			3 &{$0.4433-0.0136 i$} & {$\,\,0.4461-0.0147 i$} & {$\,\,0.4804 - 0.0024 i$} & {$\,\,0.4805 - 0.0023i$} \\
			\hline
			
		\end{tabular}
		
		\begin{tabular}{ccccc}
			
			\hline
			\\[-6pt]
			\multicolumn{1}{c}{} & \multicolumn{2}{c}{\textbf{$\Omega_0=0.001$}} & \multicolumn{2}{c}{\textbf{$\Omega_0=0.01$}} \\
			\cmidrule(rl){2-3} \cmidrule(rl){4-5}
			\textbf{$m$} & {SM} & {CFM} & {SM} & {CFM} \\
			\midrule
			-2 & {$0.2379-0.1371i $} & {$\,\,\,\,0.2362-0.1309i$}  & {$\,\,0.2496-0.0618i$}  &{$\,\,0.2487-0.0683i$}  \\
			-1 & {$0.1475-0.1106i$} & {$\,\,0.1483-0.1068i$}  & {$\,\,0.1694-0.0492i$}  & {$\,\,0.1713-0.0530i$}  \\
			1 & {$0.1811-0.1489i$} & {$\,\,0.1706-0.1292i$}  & {$\,\,0.2058-0.0614i$}  & {$\,\,0.2023-0.0668i$}  \\
			2 & {$0.2693-0.1775i$} & {$\,\,0.2619-0.1684i$}  & {$\,\,0.2903-0.0852i$}  & {$\,\,0.2974-0.0891i$} \\
			\hline
			\hline
		\end{tabular}
	\end{table*}
\begin{align}
	_2F_1(a,b;c;y)=&\frac{\Gamma(c)\Gamma(c-a-b)}{\Gamma(c-a)\Gamma(c-b)}\,_2F_1(a,b;a+b+1-c;1-y)\nonumber\\
	&+\frac{\Gamma(c)\Gamma(a+b-c)}{\Gamma(a)\Gamma(b)}(1-y)^{c-a-b}\,_2F_1(c-a,c-b;1+c-a-b;1-y),
	\label{transformation}
\end{align}
\end{widetext}
and the relation between associated Legendre polynomial function and hypergeometry function, given by
\begin{align}
	P_\beta^\mu(y)=&\frac{1}{\Gamma(1-\mu)}\left(\frac{1+y}{1-y}\right)^{\mu/2}\,\nonumber\\&\times\,_2F_1\left(-\beta,\beta+1;1-\mu;\frac{1-y}{2}\right),
\end{align}
where  $h_3^{(\text{up})}$ can be rewritten asymptotically as
\begin{align}
	h_3^{\text(up)}(r^\ast\rightarrow-\infty)={N}e^{-\mu_3r^\ast/s_3}+{M}e^{\mu_3r^\ast/s_3}
	\label{h2}
\end{align}
with
\begin{align}
	&{N}=\frac{\Gamma(1-\mu_3)\Gamma(\mu_3)}{\Gamma(-\beta_3)\Gamma(1+\beta_3)},\nonumber\\ &{M}=\frac{\Gamma(1-\mu_3)\Gamma(-\mu_3)}{\Gamma(1-\mu_3+\beta_3)\Gamma(-\mu_3-\beta_3)}.
\end{align}
One can immediately read off
\begin{align}
	&A_H^-={A}_3+{N}{B}_3,\\
	&A_H^+={M}+{B}_3.
\end{align}

When the mass shell is absent, $A_3$ and $B_3$ are found in the matching conditions with  ${A}_2=0$, ${B}_2=1$.
The spectrum then can be found by $A_H^+/A_H^-=0$ \cite{Macedo2018, Torres2022}.
Alternatively, one can start from the $i=3$ region and use \eqref{hH} to determine $A_3=1$ and $B_3=0$. The matching points are still at $\bar{r}^\ast$ and $\tilde{r}^\ast$. Using the above identity rewrites $h_1^{\text(in)}$ at $r^\ast \rightarrow \infty$. Then $A^{(+)}_{\infty}$ and $A^{(-)}_{\infty}$ can be read off from \eqref{hH}. The requirement of $A^{(-)}_{\infty}/A^{(+)}_{\infty}=0$ gives the eigenfequency.

Next, we made an extension of the continued fraction method originated from \cite{Leaver1990} for the case of  a mass shell around the black hole.
We follow the numerical treatments of calculating the QNMs and the Regge pole spectrum for a black hole surrounded by thin-shell matter in Ref.~\cite{Torres2023, Torres2023b}
The solution of the Klein-Gordon equation \eqref{KGE} can be written as a series expansion around the point $r=b$ selected to be outside the mass shell where $b>r_0+1/\alpha$ in \eqref{mass}.

After substituting it into the \eqref{KGE}, we have four-term recurrence relation given by
\begin{align}
	\alpha_ka_{k+1}+\beta_ka_k+\gamma_ka_{k-1}+\delta_ka_{k-2}=0,
	\label{re_rel}
\end{align}
for $k=2,3,\cdots$, where
\begin{align}
	\alpha_k=&-\frac{4 \left(b^2-1\right)^2 k (k+1)}{b}\, ,\\
	\beta_k=&\frac{8 \left(b^2-1\right) k \left[-i b^3 \omega +\left(b^2-3\right) k+1\right]}{b},\\
	\gamma_k=&6 b \left[4 k (2 k-3)+5\right]-\frac{5 \left[4 k (3 k-5)+9\right]}{b}\nonumber\\&-4 b \left(l^2+1\right) m^2-16 i b^2 (k-1) \omega\nonumber\\
	&+b^3 \left[-4 (k-1) k+4 m (2 l \omega +m)-1\right]\, ,\\
	\delta_k=&\frac{20 (3-2 k)^2}{b}+8 i b^2 (k-2) \omega\nonumber\\ &+4 b \left[-8 k^2+22 k+2 \left(l^2+1\right) m^2-15\right].
\end{align}
with the initial conditions 
\begin{align}
	&a_0=e^{-i\omega r^\ast(b)}H_{\omega,m}(b),\\
	&a_1=be^{-i\omega r^\ast(b)}\left[H'_{\omega,m}(r)-\frac{i\omega}{r^2-1}H_{\omega,m}(r)\right]_{r=b},
\end{align}
which can be found numerically by integrating \eqref{KGE} from the horizon $r=1$ up to $r=b>r_0+1/\alpha$.
\begin{figure}[t]
	\includegraphics[width=\linewidth]{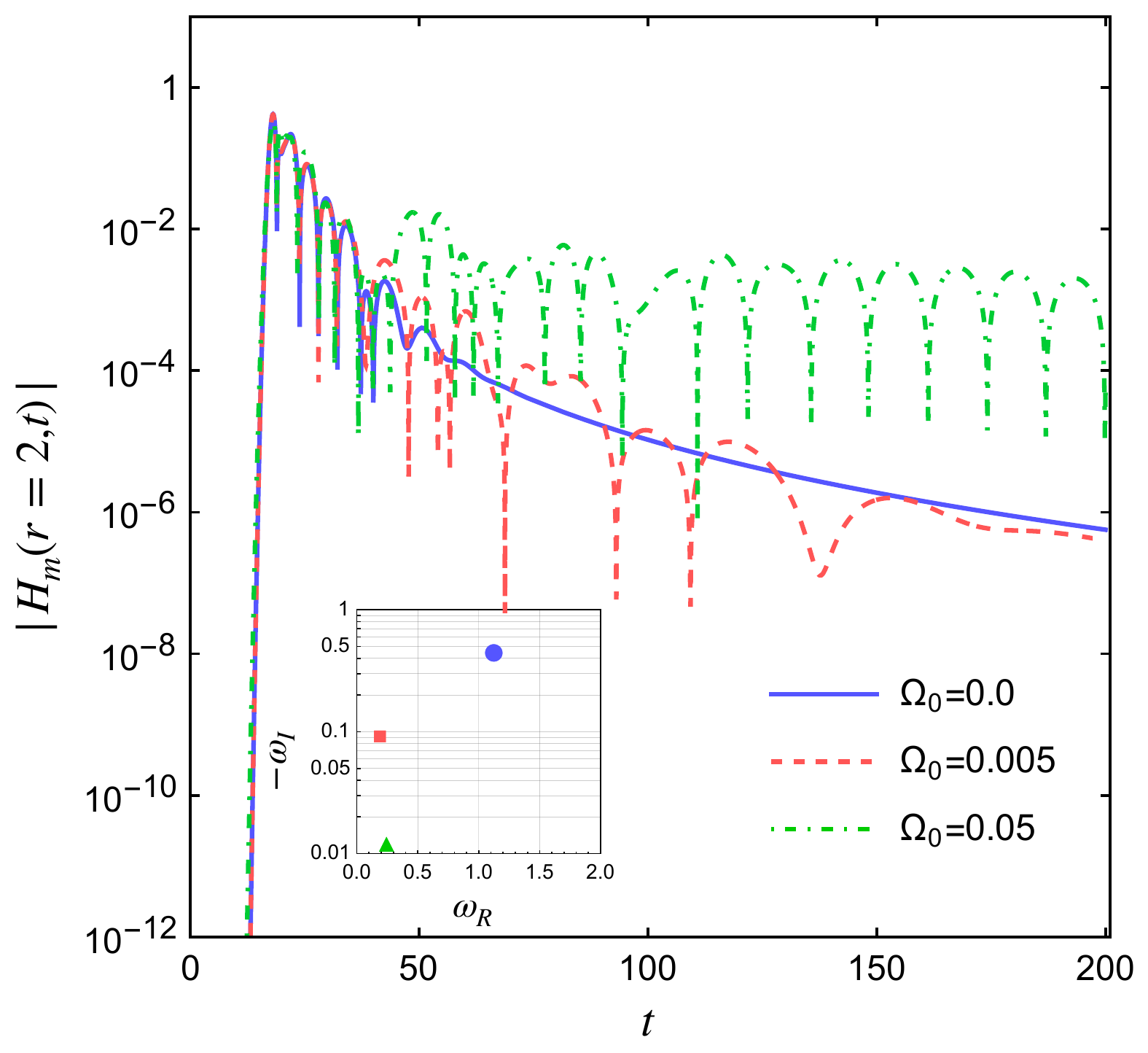}
	\caption{The absolute value of time-domain profile with varying $\Omega_0$. The inset figure shows the fundamental QNM frequency with the corresponding value of $\Omega_0$.  Other parameters are fixed as $m=1$, distance $r_0=15$ and $\alpha=1/2$.}
	\label{fig_against_small_perturbation}
\end{figure}
Furthermore, one can use Gaussian elimination to reduce \eqref{re_rel} to the three-term recurrence relation
\begin{align}
	&\alpha_0'a_1+\beta_0'a_0=0,\\
	&\alpha_k'a_{k+1}+\beta_k'a_k+\gamma_k'a_{k-1}=0, \quad k=1,2,\cdots.
\end{align}
$\alpha_k',\,\beta_k'$ and $\gamma_k'$ can be expressed in terms of $\alpha_k,\,\beta_k,\,\gamma_k$ and $\delta_k$, given by
\begin{align}
	&\alpha_0=\alpha_0', \,\beta_0'=\beta,\,\gamma_0'=\gamma_0\\
	&\alpha_1=\alpha_1', \,\beta_1'=\beta_1,\,\gamma_1'=\gamma_1\\
	&\alpha_k'=\alpha_k,\,\beta_k'=\beta_k-\alpha_{k-1}'\delta_k/\gamma_{k-1}',\nonumber\\
	\text{and}\quad&\gamma_k'=\gamma_k-\beta_{k-1}'\delta_k/\gamma_{k-1}'.
\end{align}
Regarding to the issue of convergence of series expansion (S34), we refer the study of \cite{Benhar1999} where mentioned that $a_k$ is a minimal solution to the recurrence relation when $b/2 < r_0+1/\alpha < b$ that in turn give the continued fraction as
\begin{align}
	\frac{a_1}{a_0}=-\frac{\gamma_1'}{\beta_1'-}\frac{\alpha_1'\gamma_2'}{\beta_2'-}\frac{\alpha_2'\gamma_3'}{\beta_3'-}\cdots.
\end{align}

The complex-frequency roots of the above equation corresponding to different Rabi coupling strengths are shown in Table~I, demonstrating that this modified continued fraction method  can calculate not only quasibound state but also the destabilized QNMs \cite{Cheung2022}.
\section{NUMERICALLY SOLVING THE TIME-DEPENDENT EQUATION IN (\ref{TDE}) FOR $H_m(r,t)$ }\label{appendix3}
\begin{align}
	H_{\omega,m}(r)=e^{i\omega r^\ast(r)}\sum_{k=0}^{\infty}a_k\left(1-\frac{b}{r}\right)^k.
	\label{happ}
\end{align}
In addition to the analysis in frequency domain,  we also study the time evolution of the small perturbations in the background of the presence of the spatial dependent mass term.
By numerically integrating Eq.~(\ref{TDE}) with the forth-order Runge-Kutta method, we focus on the response of the radial profile $H_m(r,t)$ at $r=2$ as a function of time.
We consider the initial conditions of the Gaussian pulse
\begin{align}
	&H(r^\ast,t=0)=\exp\left[-(r^\ast-\zeta^\ast)^2/2\sigma^2\right],\\
	&\partial_tH(r^\ast,t=0)=0,
\end{align}
where $\zeta^\ast$ is chosen to be outside the radius of  the mass term, and $\sigma$ is the width.

In Fig.~\ref{fig_against_small_perturbation}, we compare three cases with $2\Omega_0=0.0$, $0.01$ and $0.1$. Even though the spectrum has significant change for small perturbations, such as $2\Omega_0=0.01$, we find that the response waveform in the time domain shows not much difference from the unperturbed one.

\bibliography{ABH2D_TBEC.ref}
\end{document}